\documentclass[reprint,superscriptaddress,amsmath,amssymb,
aps,prl,floatfix]{revtex4-1}

\usepackage{bm}
\usepackage{float}
\usepackage{graphicx}
\usepackage[usenames,dvipsnames]{color}
\usepackage[normalem]{ulem}
\usepackage[svgnames]{xcolor}
\usepackage{bm}
\usepackage{multirow}
\usepackage{float}
\usepackage{titlesec}

\begin{document}
\preprint{APS/123-QED}


\title{Giant Seebeck effect across the field-induced metal-insulator transition of InAs}

\author{Alexandre Jaoui}
\email{alexandre.jaoui@espci.fr}
\affiliation{JEIP, USR 3573 CNRS, Coll\`ege de France, PSL Research University, 11, Place Marcelin Berthelot, 75231 Paris Cedex 05, France.}
\affiliation{Laboratoire de Physique et Etude des Mat\'eriaux (CNRS/UPMC), Ecole Sup\'erieure de Physique et de Chimie Industrielles, 10 Rue Vauquelin, 75005 Paris, France.}

\author{Gabriel Seyfarth}
\affiliation{Laboratoire National des Champs Magnétiques Intenses (LNCMI-EMFL), CNRS, UGA, UPS, INSA, Grenoble/Toulouse, France.}
\affiliation{Universit\'e Grenoble-Alpes, Grenoble, France.}

\author{Carl Willem Rischau}
\affiliation{Laboratoire de Physique et Etude des Mat\'eriaux (CNRS/UPMC), Ecole Sup\'erieure de Physique et de Chimie Industrielles, 10 Rue Vauquelin, 75005 Paris, France.}
\thanks{Present address : Department of Quantum Matter Physics (DQMP), University of Geneva, 24 Quai Ernest-Ansermet, 1211 Geneva 4, Switzerland.}

\author{Steffen Wiedmann}
\affiliation{High Field Magnet Laboratory (HFML-EMFL), Radboud University,
Toernooiveld 7, Nijmegen 6525 ED, Netherlands.}
\affiliation{Radboud University, Institute for Molecules and Materials, Nijmegen 6525 AJ, Netherlands.}

\author{Siham Benhabib}
\affiliation{Laboratoire National des Champs Magnétiques Intenses (LNCMI-EMFL), CNRS, UGA, UPS, INSA, Grenoble/Toulouse, France.}

\author{Cyril Proust}
\affiliation{Laboratoire National des Champs Magnétiques Intenses (LNCMI-EMFL), CNRS, UGA, UPS, INSA, Grenoble/Toulouse, France.}

\author{Kamran Behnia}
\affiliation{Laboratoire de Physique et Etude des Mat\'eriaux (CNRS/UPMC), Ecole Sup\'erieure de Physique et de Chimie Industrielles, 10 Rue Vauquelin, 75005 Paris, France.}

\author{Beno\^it Fauqu\'e}
\email{benoit.fauque@espci.fr}
\affiliation{JEIP, USR 3573 CNRS, Coll\`ege de France, PSL Research University, 11, Place Marcelin Berthelot, 75231 Paris Cedex 05, France.}

\begin{abstract}

\color{black} Lightly doped III-V semiconductor InAs is a dilute metal, which can be pushed beyond its extreme quantum limit upon the application of a modest magnetic field. In this regime, a Mott-Anderson metal-insulator transition, triggered by the magnetic field, leads to a depletion of carrier concentration by more than one order of magnitude. Here, we show that this transition is accompanied by a two-hundred-fold enhancement of the Seebeck coefficient which becomes as large as $11.3$mV.K$^{-1}\approx 130\frac{k_B}{e}$ at $T=8$K and $B=29$T. We find that the magnitude of this signal depends on sample dimensions and conclude that it is caused by phonon drag, resulting from a large difference between the scattering time of phonons (which are almost ballistic) and electrons (which are almost localized in the insulating state). Our results reveal a path to distinguish between  possible sources of large thermoelectric response in other low density systems pushed beyond the quantum limit.
\color{black} 
\end{abstract}

\maketitle
\section{Introduction \color{black}}
The thermoelectrical properties of low carrier density metals are of fundamental and technological interest. Due to their small Fermi temperatures, their diffusive Seebeck effects ($S_{xx}$) can be large and, as such, used to develop high performance thermoelectric devices \cite{NolasBook}. At low temperature, their thermoelectrical response is also a fine probe of their fundamental electronic properties, in particular in the presence of a magnetic field \cite{Behniabook}. As an example, the large quantum oscillations observed in the Nernst effect ($S_{xy}$) of semi-metals \cite{Behnia2007,Zhu2010,Zhu2011} or doped semi-conductors \cite{Tieke1996,liang2013,Zhang2020} have been used to map out their Fermi surfaces and to reveal the Dirac/Weyl nature of the electronic spectrum of Bi\cite{Zhu2011}, Pb$_{1-x}$Sn$_x$Se \cite{liang2013} or ZrTe$_5$ \cite{Zhang2020}.
For most dilute metals, a magnetic field of a few Tesla (labelled $B_{QL}$) is enough to confine all the charge carriers in the lowest Landau level (LLL), the so-called quantum limit. At low temperature, this is concomitant with an increase of $S_{xx}$ (and $S_{xy}$) \cite{Puri1964,liang2013,fauque2013_bi2se3,Zhang2020}. This increase can be found at higher temperatures, as illustrated by the large Seebeck effect observed in the quantum limit regime of the Weyl semi-metal TaP \cite{han2019discovery} or in the fractional quantum Hall regime of 2DEGs \cite{Fletcher1986}. In these systems, the amplitude of $S_{xx}$ is much larger than $\frac{k_B}{e}$, the natural thermoelectric scale of the diffusive response. This surprisingly large $S_{xx}$ can be either the result of an unbounded diffusive thermoelectric power specific to nodal metals \cite{Skinnereaat2018} or to the coupling of electrons with the phonon bath. In the latter case, the so-called phonon drag, the amplitude of $S_{xx}$ is dictated by the momentum transfer between the electron and phonon baths and can be much larger than $\frac{k_B}{e}$ \cite{Herring1954}. The phonon drag effect is well known to enhance $S_{xx}$ in doped semi-conductors or metals at zero magnetic field but also in the quantum limit regime \cite{Puri1964, Jay-Gerin1975}.\\
Here, we present a study of the electrical and thermoelectrical properties of InAs, a bulk narrow-direct-gap semiconductor, beyond its quantum limit and up to a so-far unexplored range of temperature and magnetic field. Our findings show that the field induces a metal-insulator transition (MIT) that is accompanied by a giant peak in the Seebeck effect, as large as $S_{xx}$=11.3mV.K$^{-1}$ at $T$=8K. Based on a  study of the thermal response as a function of sample dimensions, we argue that this giant Seebeck effect results from a phonon drag effect. Moreover, at the highest magnetic field, and in contrast with the well-documented case of lightly doped InSb, we show that the electronic properties of InAs are characterised by a residual conductance and thermopower originating from poorly mobile bulk carriers and highly mobile carriers on the surface.

\begin{figure}
\begin{center}
\makebox{\includegraphics[width=0.5\textwidth]{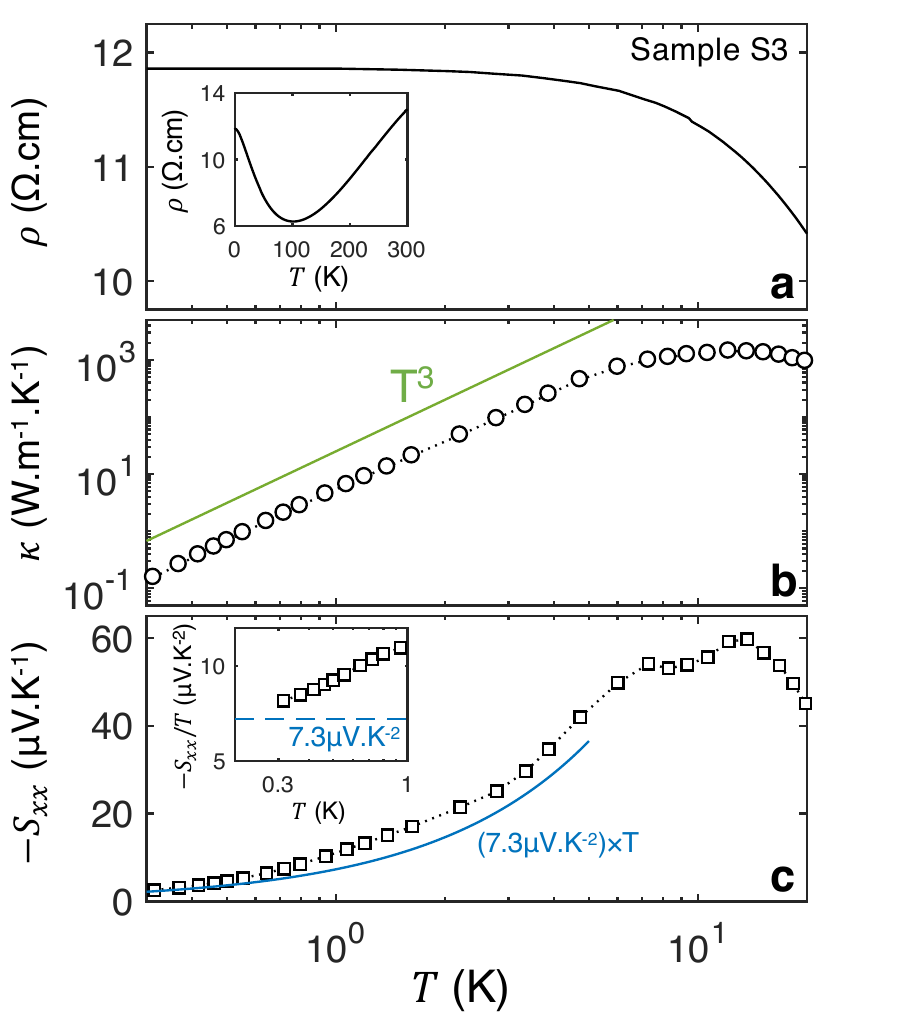}}
    \caption{\textbf{a)} Resistivity ($\rho$) \textbf{b)} Thermal conductivity ($\kappa$) and \textbf{c)} $-S_{xx}$ of InAs sample S3 as functions of the temperature at $B=0$T (see \cite{SM} for the sample descriptions). Inset in a) shows $\rho$ up to $T=300$K. The solid green line in b) represents a $T^3$ power law characteristic of phonon contribution in the ballistic regime. The inset in c) shows $-\frac{S_{xx}}{T}$ plotted as a function of the temperature in the sub-Kelvin region. The blue dashed line corresponds to the value $\frac{S_{xx}}{T}=7.3\mu$V.K$^{-2}$ which is expected in the diffusive regime of a degenerate electron gas with $T_F=100$K (see \cite{SM}). The diffusive contribution is further emphasized as a blue line in c).}
   \label{fig1}
   \end{center}
\end{figure}

\section{A dilute metal at zero magnetic field}

We show in Fig.\ref{fig1} the temperature dependence of the resistivity ($\rho_{xx}$), the Seebeck coefficient ($S_{xx}$) and the thermal conductivity ($\kappa$) of an $n$-type InAs with a Hall carrier density $n_H=2.0\times10^{16}$cm$^{-3}$. The Fermi surface of InAs, studied by low magnetic field quantum oscillation measurements \cite{SM}, is formed by a single spherical pocket located at the $\Gamma$-point of the Brillouin zone with a carrier density $n_{SdH}=1.6\times10^{16}$cm$^{-3}$ ($T_F=100$K) and mass carrier $m^*=0.023m_0$. From room temperature down to $T_F$, $\rho_{xx}$ is metallic and the non-degenerate electrons are mainly scattered by phonons. Below $T_F$, $\rho_{xx}$ increases with decreasing temperature as electrons become more and more degenerate with a mobility limited by ionized-impurity scattering \cite{Rode1970}. Below $T=20$K, $\rho_{xx}$ is constant with a residual resistance of $\rho_0=12$m$\Omega$.cm and a Hall mobility $\mu_H=24000$cm$^{2}$.V$^{-1}$.s$^{-1}$. The electronic contribution ($\kappa_{el}$) to $\kappa$ is given by the Wiedemann Franz law $\kappa_{el}=\frac{L_0T}{\rho_0}= 0.19$mW.K$^{-2}$.m$^{-1}$. It is much smaller than the $\kappa$ shown in Fig.\ref{fig1}.b and points to a purely phononic origin to $\kappa$. Below $T=6$K, $\kappa$ scales with $T^3$. The phonon mean free path, $\ell_{ph}$, can be estimated through the kinetic formula $\kappa$ = $\frac{1}{3}C_{ph}v_{ph}\ell_{ph}$, where $C_{ph} = \beta_{ph} T^{3}$ is the specific heat associated with phonons at low temperature with $\beta_{ph}$ = 3.68J.K$^{-4}$.m$^{-3}$ \cite{Cetas1978} and $v_{ph}$ is the sound velocity $v_{ph}=2.5$km.s$^{-1}$ \cite{Guillou1972}. The deduced $\ell_{ph}$ is of the order of sample thickness ($e$) : phonons are in the ballistic regime. Similarly to $\kappa$, $S_{xx}$ peaks around $T=10$K. In the zero temperature limit $\frac{S_{xx}}{T}$ saturates to a value of $7.8\mu$V.K$^{-2}$ which is in quantitative agreement with the expected value for the diffusive response of a degenerate semiconductor in the ionized impurity scattering regime \cite{SM}. The finite-temperature extra contribution to $S_{xx}$ comes from the phonon-drag effect. In summary, at zero magnetic field, InAs is a dilute metal with one electron per $10^6$ atoms. The mobility of these carriers does not evolve much with cooling. Yet, it is high enough to allow the observation of quantum oscillations. Thermal transport is dominated by phonons, which become ballistic below $T=10$K while the thermoelectric response is purely diffusive in the zero temperature regime with a modest phonon drag component at finite temperature.

\color{black}

\begin{figure}
\centering
\makebox{\includegraphics[width=0.5\textwidth]{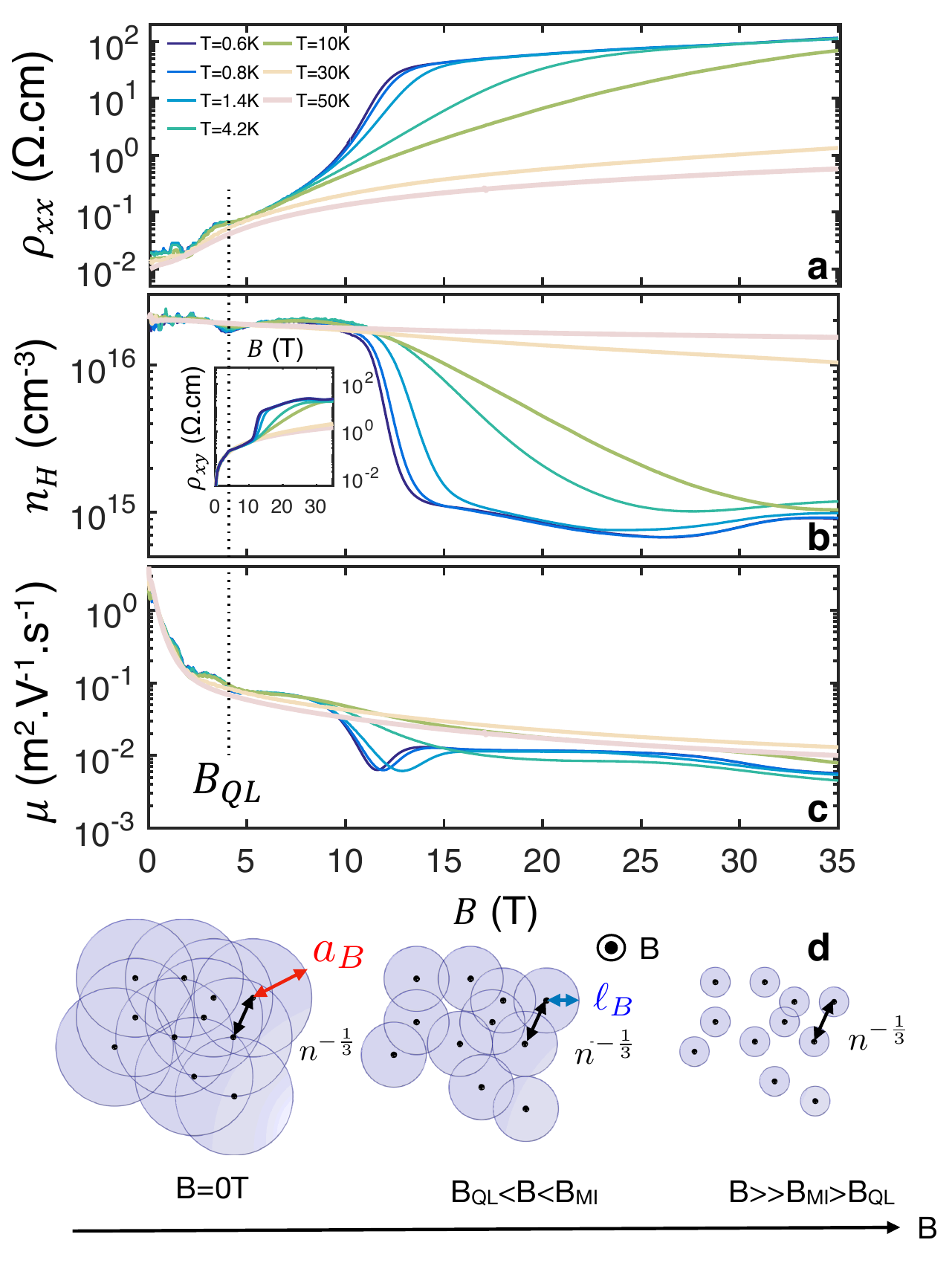}}
    \caption{\textbf{a)} Transverse ($\vec{j}\perp \vec{B}$) magnetoresistivity of sample S1 for various temperatures up to $B=35$T. The evolution of the Hall carrier concentration $n_H$ with the magnetic field, for the same sample and temperatures, is featured in \textbf{b)} alongside the electronic mobility $\mu$ in \textbf{c)}. Inset of c) shows the Hall resistivity of the sample. The vertical dotted line marks the quantum limit at $B_{QL}$. 
    \textbf{d)} Sketch of the magnetic freeze-out effect : as the magnetic field increases the effective Bohr radius $a^{*}_{B}$=$(a_{B,\perp}^2a_{B,\parallel})^{\frac{1}{3}}$ (see the text for the definition) is decreasing. When the overlap between the wave function is sufficiently reduced, a metal-insulator transition is expected to occur at $B_{MI}$ when the condition  $n^{1/3}(a^{*}_{B}(B=B_{MI}))=\delta$ is satisfied where $\delta\approx0.3-0.4$.}
    \label{fig2}
\end{figure}

\begin{figure*}
\centering
\makebox{\includegraphics[width=0.8\textwidth]{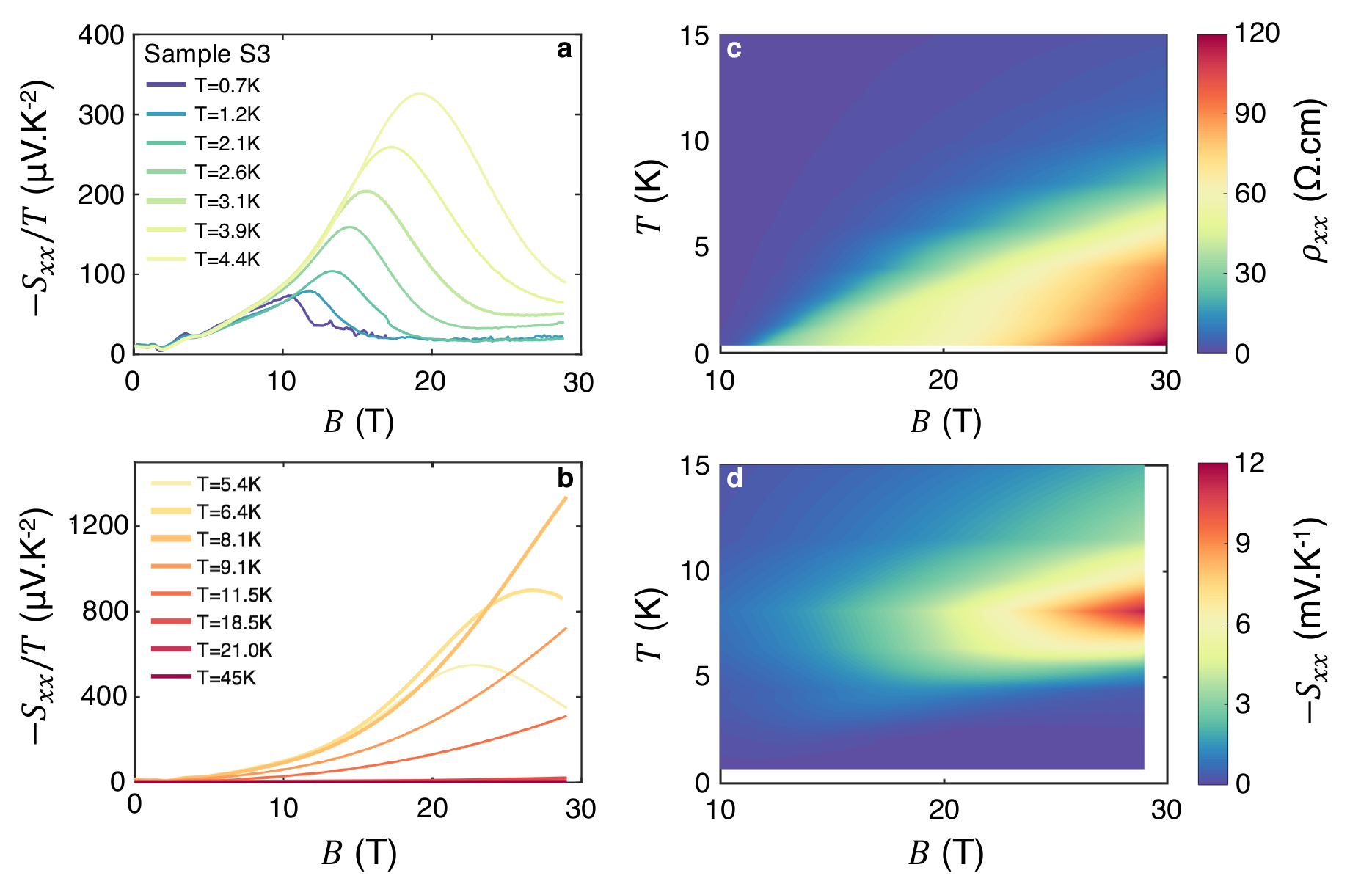}}
    \caption{ \textbf{a)} $-\frac{S_{xx}}{T}$ of sample S3 as a function of the magnetic field from $T=0.7$K to $4.4$K and \textbf{f)} from $T=5.4$K to $45$K. \textbf{c)} and \textbf{d)} Temperature-magnetic field color map of the transverse resistivity $\rho_{xx}$ and $-S_{xx}$ of sample S3.}
    \label{fig4}
\end{figure*}

\section{Field-induced metal-insulator transition in InAs}
\begin{figure}
\begin{center}
\makebox{\includegraphics[width=0.5\textwidth]{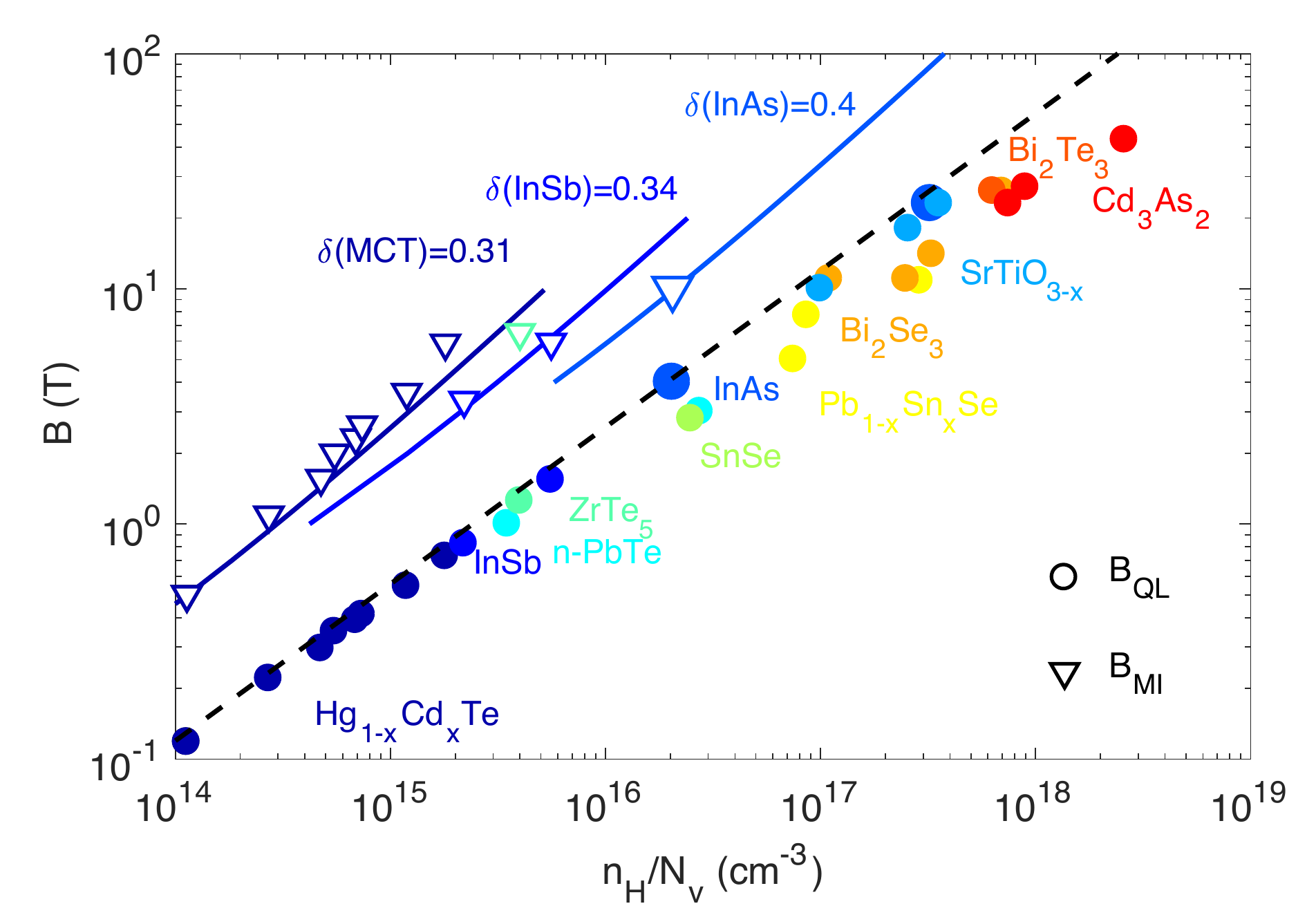}}
    \caption{Observed metal-insulator transition fields (taken as the drop in the Hall carrier density), $B_{MI}$ (open triangles), and the positions of the last maximum in the resistivity or the Nernst effect, $B_{QL}$ (open circles) plotted as functions of the
    carrier densities of doped semi-conductors. The points are extracted from references\cite{shayegan1988,Oswald1989,Assaf2017,liang2013,Wang2018,Kohler1975,Analytis2010,Rischau2012,Tang2019,Xiang2015,Narayanan2015,Zhao2015,Bhattacharya2016} (see \cite{SM} for further details).  The dashed black line is the predicted $B_{QL}=\frac{\hbar(3\pi^{2}n)^{\frac{2}{3}}}{e(1+\mid M \mid)}$ for an isotropic parabolic dispersion with the $g$-factor of InAs ($g(InAs)=15$ $M=0.13$). The solid curves represent the metal-insulator transition field vs density calculated using Eq. \ref{Eq:MOT} with $\delta=0.31$(Hg$_{0.8}$Cd$_2$Te), $0.34$(InSb) and $0.4$(InAs).}
   \label{figBQL}
   \end{center}
\end{figure}

Let us now discuss the evolution in field of the electrical and thermoelectrical properties of InAs which are shown respectively on Figs.\ref{fig2} and \ref{fig4}. At low magnetic fields, $\rho_{xx}$, $n_H$ and $S_{xx}$ display quantum oscillations. The last quantum oscillation occurs at $B_{QL}=4.1$T \cite{SM}. Above this field all the carriers are confined in the lowest Landau level (LLL). Up to $B=10$T, $\rho_{xx}$ and $S_{xx}$ increase while $n_H$ remains constant. From $B=10$T to $B=15$T, $\rho_{xx}$ increases by about two orders of magnitude while $n_H$ drops by a factor close to 20 and $S_{xx}$ by a factor of three. This marks the entrance in the magnetic freeze-out regime, in good agreement with early measurements \cite{kaufman1970,kadri1985}. This regime has been thoroughly studied in the lightly doped narrow gap semi-conductors InSb and Hg$_{1-x}$Cd$_x$Te (n$_H$=10$^{-14}$-10$^{-15}$cm$^{-3}$) \cite{shayegan1988,Aronzon1990}. It was shown that the magnetic field induces a MIT ascribed as a magnetic field assisted Mott-Anderson transition in lightly doped semiconductors (see \cite{Aronzon1990} for a review). A sketch of this effect is shown in Fig.\ref{fig2}.d. For $B>B_{QL}$, the in-plane electronic wave extension is equal to $a_{\perp}=2\ell_B$ with $\ell_B=\sqrt(\frac{\hbar}{eB})$ and shrinks with the magnetic field. Parallel to the magnetic field the characteristic spatial extension is given by $a_{B,\parallel}=\frac{a_B}{\log(\gamma)}$ where $\gamma=(\frac{a_B}{l_B})^2$ \cite{yafet1956,Shklovskii1984}. Once the overlap between the wave functions of electrons is sufficiently reduced, a MIT is expected to occur at $B=B_{MI}$, i.e. when :
\begin{equation}
n^{1/3}(a_{B,\perp}^2a_{B,\parallel})^{\frac{1}{3}}=\delta
\label{Eq:MOT}
\end{equation}
with $\delta=0.3-0.4$ for InSb and Hg$_{1-x}$Cd$_x$Te. The carrier dependence of $B_{QL}$ and $B_{MI}$ for these two systems are shown in Fig.\ref{figBQL}. Using the drop of $n_H$ in the zero temperature limit \cite{Rosenbaum1985,shayegan1988} we find that $B_{MI}=10.1\pm0.2$T \cite{SM} which is well captured by Eq.\ref{Eq:MOT} assuming a $\delta=0.4$ as illustrated in Fig.\ref{figBQL}. In the $k$-space this transition corresponds to a transfer of electrons from the LLL to a shallow band (formed by the localized electrons) located at an energy below the LLL \cite{yafet1956}.

\begin{figure*}
\centering
\makebox{\includegraphics[width=1\textwidth]{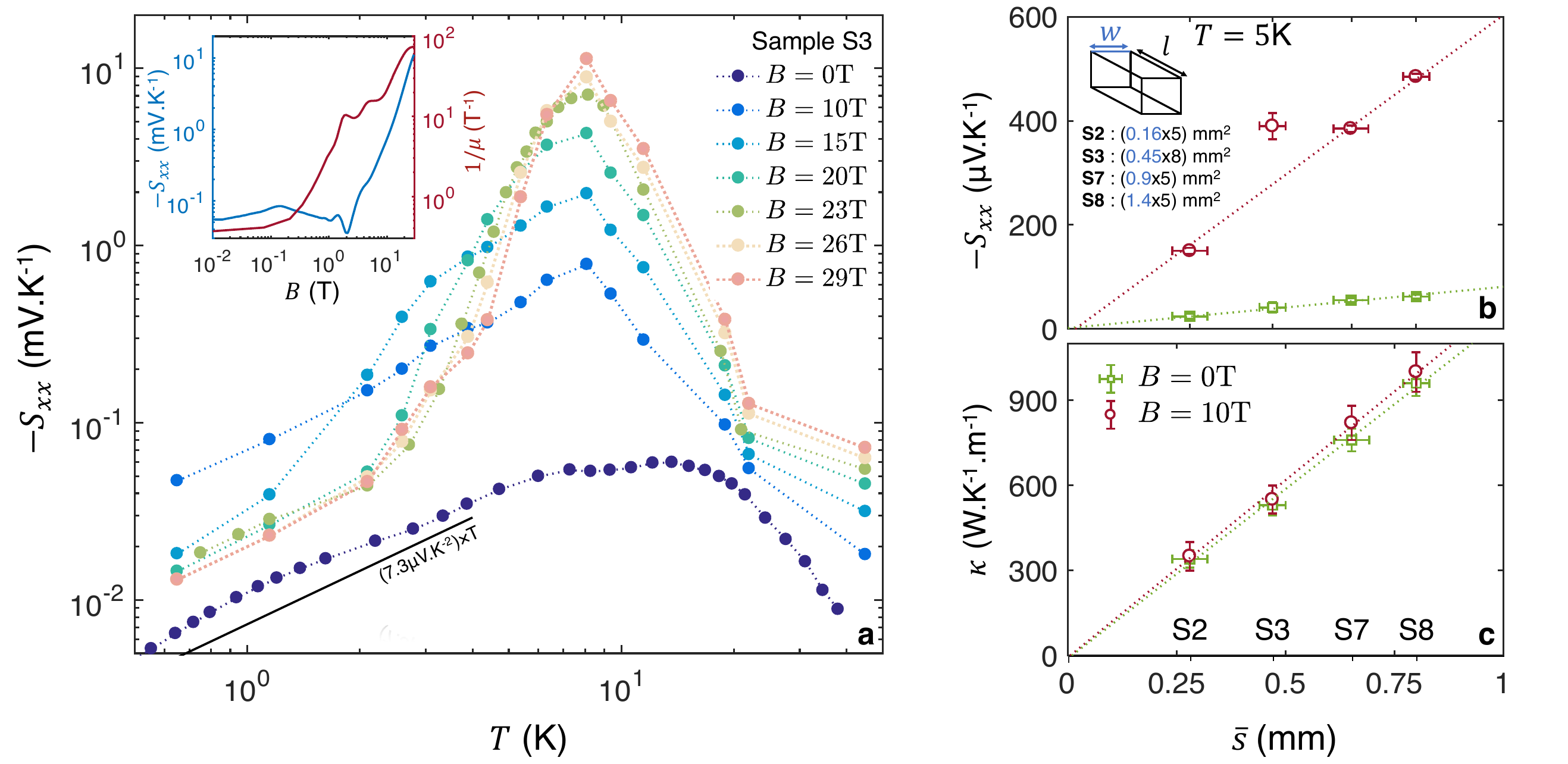}}
    \caption{\textbf{a)} $-S_{xx}$ (sample S3) as a function of temperature for various magnetic fields. The solid black line shows the thermopower value expected in the diffusive regime of a degenerate electron gas with $T_F=100$K. The inset shows a comparison of the field dependence of $-S_{xx}$ (left axis in blue) and $1/\mu_H$ (right axis in red) at $T=8$K. Both quantity are amplified by about a factor $200$ between $0$T and $30$T. \textbf{b)} and \textbf{c)} size dependence of $-S_{xx}$ and $\kappa_{xx}$ at $B=0$T (green circles) and $B=10$T (red squares) at $T=5$K. The inset describes the geometry of the samples and $\overline{s}=\sqrt($width$\times$thickness)-factor used to describe the average cross-section of the samples.}
    \label{fig5_}
\end{figure*}

Above $B=15$T, a novel regime is identified where $\rho_{xx}$ varies almost linearly with the magnetic field and concomitant with a saturating $n_H$=10$^{15}$cm$^{-3}$ and $\frac{S_{xx}}{T}$=$-21\mu$V.K$^{-2}$ at the lowest temperature. As a function of the temperature, $\rho_{xx}$ first displays an activated behavior followed by a saturation at low temperature (see \cite{SM}). The deduced gap, $\Delta$, is equal to $2$meV at $B=30$T and increases with the magnetic field up to $B=50$T \cite{SM}. As the temperature is increased, the transition shifts to higher magnetic field, becomes broader and vanishes above $T=30$K in the electrical response. Likewise, the peak in $S_{xx}$ shifts to higher magnetic field. However, surprisingly, its amplitude increases. This striking observation is better appreciated by comparing the two color-maps of $\rho_{xx}$ and $S_{xx}$ which are shown respectively in Figs.\ref{fig4}.c and d. While $\rho_{xx}$ is maximal at the lowest temperature, $S_{xx}$ is maximal at around $T=8$K and that for all magnetic fields as shown in Fig.\ref{fig5_}.a. At $B=29$T, $S_{xx}(T=8$K$)$ is as large as $11.3$mV.K$^{-1}$ which is about two hundred times larger than the zero-field thermopower and comparable with the "colossal" thermopower observed in ultra-low carrier density Ge ($n_{H}<10^{13}$cm$^{-3}$) where $S_{xx}(T=10$K$)\approx 10-30$mV.K$^{-1}$ \cite{Inyushkin2003}, in the strongly correlated semi-conductors FeSb$_2$ ($n_H\approx 1 \times10^{15}$cm$^{-3}$) where $S_{xx}(T=10$K$)\approx10-30$mV.K$^{-1}$ \cite{Bentien2007,Takahashi2016} or in the fractional quantum Hall regime of 2DEGs where $S_{xx}$($T=5$K) reaches $50$mV.K$^{-1}$ \cite{Fletcher1986}. Let's now discuss on the origin of the two intriguing properties identified above $B_{MI}$: the saturating magnetoresistance and the giant Seebeck response. 

\section{Bulk vs surface state contribution to electric and thermoelectrical properties}

The residual conductivity and carrier density at low temperature and high magnetic field contrasts with the insulating behavior of lightly doped InSb \cite{shayegan1988}. A key difference between both systems is the length scale of the fluctuations of the impurity potential. In highly doped semiconductors, large-scale fluctuations affect the density of state, in particular in its quantum limit regime \cite{dyakonov1969}. Scanning tunneling microscopy (STM) measurements have revealed spatial fluctuations of the LLL in InAs on the energy scale of $\Gamma=3–4$meV which result in a broadening of the shallow band and a tail in the density of state of the LLL \cite{Morgenstern2000}. Such broadening manifests itself in the electrical transport properties by a gap ($\Delta$) value much lower than the theoretical predictions \cite{kaufman1970} and a residual carrier concentration down to the lowest temperature well above $B_{MI}$ as far as $\Gamma\approx\Delta$ \cite{dyakonov1969}. With a residual carrier density of $n_{H,B>B_{MI}}=1\times10^{15}$cm$^{-3}$ and a resistivity $\rho_{xx}\approx100\Omega$.cm, these residual bulk electrons have a low mobility $\mu_{H,B>B_{MI}}=60$cm$^{2}$.V$^{-1}$.s$^{-1}$. Such poorly mobile carriers can be shunted by the conductance of the surface as it is the case in the magnetic freeze-out regime of Hg$_{1-x}$Cd$_x$Te \cite{Mullin_1984,Antcliffe1970}. Magneto-transport, magneto-optical measurements \cite{Tsui1970,Reisinger1981} and ARPES measurements \cite{olsson1996,King2010} on InAs have well documented the existence of an accumulation layer of carrier density $n_S=1\times10^{12}$cm$^{-2}$($E_F\approx100$meV) of mobility $\mu_{H,S}\approx5000$cm$^{2}$.V$^{-1}$.s$^{-1}$. For a sample thickness, $e$, the ratio of conductance from the bulk ($\sigma_b$) and the surface ($\sigma_s$) is given by: $\frac{\sigma_b}{\sigma_s}=\frac{en_{H,B>B_{MI}}\mu_{H,B>B_{MI}}}{n_{S}\mu_{H,s}}$. With the numbers given above, $\frac{\sigma_b}{\sigma_s}\approx 1$ : both contributions are of the same order of magnitude. This is supported by the amplitude of the thermopower at low temperature for $B>B_{MI}$. In the presence of bulk and surface contributions, $S_{xx}$ is given by the sum of the bulk and surface thermopower (labelled $S_{xx,B}$ and $S_{xx,S}$) balanced by their relative contribution to the total conductivity: $S_{xx}=\frac{\sigma_{B}S_{xx,B}+\sigma_{S}S_{xx,S}}{\sigma_B+\sigma_S}$. With a Fermi temperature of the surface states ($T_{F}\approx900$K\cite{Reisinger1981}) much larger than the Fermi temperature of the residual bulk state ($T_{F,B>B_{MI}}=18$K) $S_{xx,S}<<S_{xx,B}$ with $\frac{S_{xx,B}}{T}$ expected to be $-46\mu$V.K$^{-2}$ in the diffusive regime of ionised impurity scattering. With $\sigma_B\approx\sigma_S$, $\frac{S_{xx}}{T}\approx\frac{S_{xx,B}}{2T}=-23\mu$V.K$^{-2}$ in good agreement with the residual measured thermopower. Contributions from the surface states is further supported by the sample dependence of the low-temperature-high-field value of $\rho_{xx}$ as discussed in \cite{SM}. Therefore, deep inside its quantum limit, the electrical transport properties of InAs reveal two types of contributions : a first from low mobility bulk electrons and a second from highly mobile electrons on the surface.

\section{Quantifying the phonon drag contribution}
 
The field dependence of $S_{xx}$ is qualitatively well captured by the Mott relation \cite{Ziman}($\frac{S_{xx}}{T}=-\frac{\pi^2}{3}\frac{k_B^2}{e}\frac{\partial \ln(\sigma(\epsilon))}{\partial \epsilon}_{\epsilon=\epsilon_F}$): this is in the region where $\rho_{xx}$ and $\rho_{xy}$ (and thus $\sigma_{xx}$) vary the most with both the magnetic field and temperature so that $S_{xx}$ is the largest. As a function of the magnetic field, it happens close to $B_{MI}$ leading to a peak in the field dependence of $S_{xx}$. However this relation fails to explain quantitatively the temperature evolution and the amplitude of the peak. As the temperature increases, the transition becomes broader and the amplitude of the peak is expected to vanish. In the activation regime $S_{xx}\propto\frac{k_B}{e}\frac{\Delta}{T}$. At $T=8$K and $B=30$T ($\Delta=2$meV), $S_{xx}$($T=8$K) is at most  $3\frac{k_B}{e}=250\mu$V.K$^{-1}$ fifty times smaller than the measured value. Therefore another source of entropy has to be invoked such as the phonon bath.

At zero magnetic field the phonon-drag effect is known to enhance the thermoelectrical response of doped semiconductors or metals. The phonon drag picture conceived by Herring~\cite{Herring1954} quantifies the additional contribution to the Peltier coefficient, $\Pi$ by the thermal current carried by phonons. The Peltier coefficient, the ratio of heat current to charge current, is linked to the Seebeck coefficient through the Kelvin relation. According to Herring, the phonon drag contribution to the Peltier coefficient is:
\begin{equation}
    \Pi_{ph}= \pm \Lambda m^*v_{s}^2 \frac{\tau_{ph}}{\tau_{e}}
    \label{Eq:PeltierDrag}
\end{equation} 
Here, $\Lambda<1$ quantifies the momentum exchange rate between phonons and electrons~\cite{Herring1954}, $\tau_{ph}$ and $\tau_{e}$ are phonon and electron scattering rates, $m^*$ is the effective mass and $v_{ph}$ is the sound velocity. One can see that phonon drag requires a finite $\Lambda$ and is boosted by a large $\frac{\tau_{ph}}{\tau_{e}}$ and/or a large effective mass (like in FeSb$_2$ \cite{Takahashi2016}). Using the Kelvin relation, Eq.\ref{Eq:PeltierDrag} leads to: 

\begin{equation}
    S_{ph}= \pm \frac{k_B}{e}\Lambda \frac{m^*v_{ph}^2}{k_BT} \frac{\tau_{ph}}{\tau_{e}}= \pm \Lambda \frac{\ell_{ph}v_{ph}}{\mu_{H}T}
    \label{Eq:SeebeckDrag}
\end{equation} 

Thus, a large Seebeck response in units of $\frac{k_B}{e}$ is possible thanks to phonon drag. It requires $\tau_{ph}\gg\tau_{e}$ and a finite $\Lambda$. Herring showed that in intrinsic semiconductors, such as Si and Ge, the large $\frac{\tau_{ph}}{\tau_{e}}$ ratio provides a key to understanding the large magnitude of the Seebeck response at cryogenic temperatures~\cite{Geballe1954,Geballe1955}.

Let us discuss how the magnetic field  squeezes $\tau_{e}$, leaves $\tau_{ph}$ unaffected and thus boosts their ratio. In contrast to the diffusive response, $S_{ph}$ scales with the sample length, the inverse of the mobility and can exceed by far $\frac{k_B}{e}$. A crucial test of Eq.\ref{Eq:SeebeckDrag} comes from the size and mobility dependence of $S_{xx}$. As shown in Fig.\ref{fig5_}.b) and c), both $S_{xx}$ and $\kappa_{xx}$ are size dependent and scale with the sample cross-section ($\overline{s}\propto l_{ph}$). However, the slope of $\kappa_{xx}$ vs. $\overline{s}$ is independent of the magnetic field (since it only depends on $l_{ph}$), but the slope of $S_{xx}$ vs. $\overline{s}$ increases with increasing magnetic field due to the reduction of $\mu_H$. As expected from Eq.\ref{Eq:SeebeckDrag} and illustrated in the inset of Fig.\ref{fig5_}.a), the changes induced by magnetic field in $-S_{xx}$ and in $\mu^{-1}_H$ are comparable. At T= 8K, between $B=0$T and 29T, $-S_{xx}(T=8$K$)$ is amplified by a factor of $202$ and $\mu_H$ decreases by a factor of 196. Using Eq.\ref{Eq:SeebeckDrag} at high magnetic field, we find that $\Lambda \ll 1$. This is not a surprise, given the temperature dependence of $\rho_{xx}$ which shows that  below $T=100$K, electrons are mostly scattered by ionized impurities and not by phonons.
 
The phonon drag picture therefore provides a quantitative agreement of the giant field-induced Seebeck effect in InAs. We note that below $T=8$K, $S_{xx}$ peaks in magnetic field while $S_{ph}$ is expected to be the largest at the highest magnetic field (where $\mu_H$ is the lowest). This implies either a shunt of $S_{ph}$ by the relatively small thermoelectrical response of the surface states or a more elaborate bulk phonon drag picture that would be maximum close to $B_{MI}$, where the bulk shallow band is partially filled as it has been proposed in FeSb$_2$ \cite{Bentien2007}. Interestingly FeSb$_2$ and InAs above $B_{MI}$ share in common the same activation gap (of the order of a few meV), the same carrier density ($10^{15}$cm$^{-3}$) and the presence of bulk in-gap states. However with resistivity values two orders of magnitude larger in InAs than in FeSb$_2$, the power factor of InAs is only $10\mu$W.K$^{-2}$.cm$^{-1}$ (two order of magnitude smaller than in FeSb$_2$).
 
While a purely electronic mechanism has been recently proposed to give rise to an unbounded thermopower in Dirac/Weyl semimetals in their quantum limit regime \cite{Skinnereaat2018} our results show that the phonon drag effect is another road to boost the diffusive response of low carrier density metals across their field-induced MIT. Up to now this transition has been studied in a limited number of cases and the thermoelectric properties of dilute metals remains vastly unexplored. As illustrated in Fig.\ref{figBQL}, a large class of materials (ranging from well known doped semiconductors to new topological materials) remains to be studied, in particular at higher doping (and therefore at high magnetic field) where larger $\Lambda$ can be attained, favoring even larger $S_{ph}$. 

\section{Acknowledgments}
We thanks useful discussion with A. Akrap, R. Daou, B. Skinner and J. Tomczak. This work is supported  by JEIP-Coll\`{e}ge de France and by the Agence Nationale de la Recherche  (ANR-18-CE92-0020-01, ANR-19-CE30-0014-04).

\bibliography{biblio}
\end{document}